\documentclass[pra,twocolumn,amsfonts,amssymb,amsmath,floatfix,floats,a4paper]{revtex4-1}

\usepackage{graphicx}
\begin{document}

\title{Modification of ``Counterfactual  communication protocols'' which eliminates weak
particle traces }

\author{Yakir Aharonov}
\affiliation{ Raymond and Beverly Sackler School of Physics and Astronomy\\
 Tel-Aviv University, Tel-Aviv 69978, Israel}
\affiliation{Schmid College of Science, Chapman University, Orange CA 92866, USA}
\author{Lev Vaidman}
\affiliation{Raymond and Beverly Sackler School of Physics and Astronomy\\
 Tel-Aviv University, Tel-Aviv 69978, Israel}

\begin{abstract}
 Possibility to communicate between spatially separated regions, without  even a single photon passing between the two parties, is an amazing quantum phenomenon. The possibility of transmitting one value of a bit in such a way, the interaction-free measurement, was known for quarter of a century. The protocols of full communication, including transmitting unknown  quantum states were proposed only few years ago, but it was shown that in all these protocols the particle was leaving a weak trace in the transmission channel, the trace  larger than the trace left by a single particle passing through the channel. This made the claim of counterfactuality of these protocols at best controversial. However, a simple modification of these recent protocols  eliminates the trace in the transmission channel making all these protocols counterfactual.
      \end{abstract}
\maketitle

\section{Introduction}

The beginning of counterfactual communication was when Penrose \cite{Penrose} coined the term ``counterfactuals'' for describing quantum interaction-free measurements (IFM) \cite{IFM}. The idea was developed  to  counterfactual cryptography  \cite{Noh}, to counterfactual computation \cite{Joz},   to contractual computation for all outcomes \cite{Ho06}, and then to counterfactual communication \cite{Salih}. More research about counterfactual protocols was done  \cite{CFInt3,CFInt4,CFInt5,CFInt6,Li-invi,1man,CFInt7,CFInt8,Shukur,Guo}, and even  a new kind of teleportation  \cite{tele} which required no prior entanglement, no classical channel and no particles traveling between the parties was proposed \cite{SalihQT,Li}. One of us, LV, although being a co-author of the original work \cite{IFM} criticized many counterfactual protocols as being not counterfactual  \cite{V07,V14,V14R,count,Mycom}. He showed that while the original IFM and all other protocols including counterfactual cryptography  relying on communication of only one value of a bit were indeed counterfactual, the protocols for full  communication and  computation with two values of a bit  were  not counterfactual.  Here we present a simple modification of theses protocols which makes them  counterfactual.

The basic definition of counterfactual communication is communication without particles in the transmission channel.
It is enough that (counterfactually) the particles could have been in the channel, and/or they were in the channel in runs which were discarded in the communication protocol. The controversy about counterfactuality of the protocols was about definition of ``particles being in the transmission channel''. The authors considering the protocols as counterfactual relied on classical reasoning: if the particle could not pass through the channel, it was not there. Vaidman claimed that one cannot use classical argumentation for discussing quantum particles and suggested the weak trace definition. When a particle passes through a channel it always slightly changes the quantum state of the channel, it leaves a weak trace. There is a nonzero local coupling of any particle in any channel, but the coupling can be made essentially arbitrary small. If in the communication protocol the trace left in the transmission channel is of the order (or larger) than the trace  left by a passing single particle, then, by definition \cite{past}   the particle was in the channel, and thus the protocol is not counterfactual. If the protocol leaves a trace  smaller by an arbitrary large factor than the trace of a single particle passing through the same channel, we define that the particle was not there, i.e., that the protocol is counterfactual. The suggested protocols \cite{Ho06,Salih,SalihQT,Li}  leave traces larger than the trace of a single particle passing through the transmission channel, so they are not counterfactual in this sense.

The two-state vector formalism (TSVF) \cite{AV90} provides a very simple way to find out when the trace is present: If there is an overlap of the forward and backward evolving states in the channel, then  local interactions operators in the channel do not vanish and, therefore, the particle leaves a weak  trace in the channel. Thus, by definition, the particle was present in the channel, i.e. the protocol is not counterfactual.

\begin{figure}
\begin{center}
 \includegraphics[width=8.2cm]{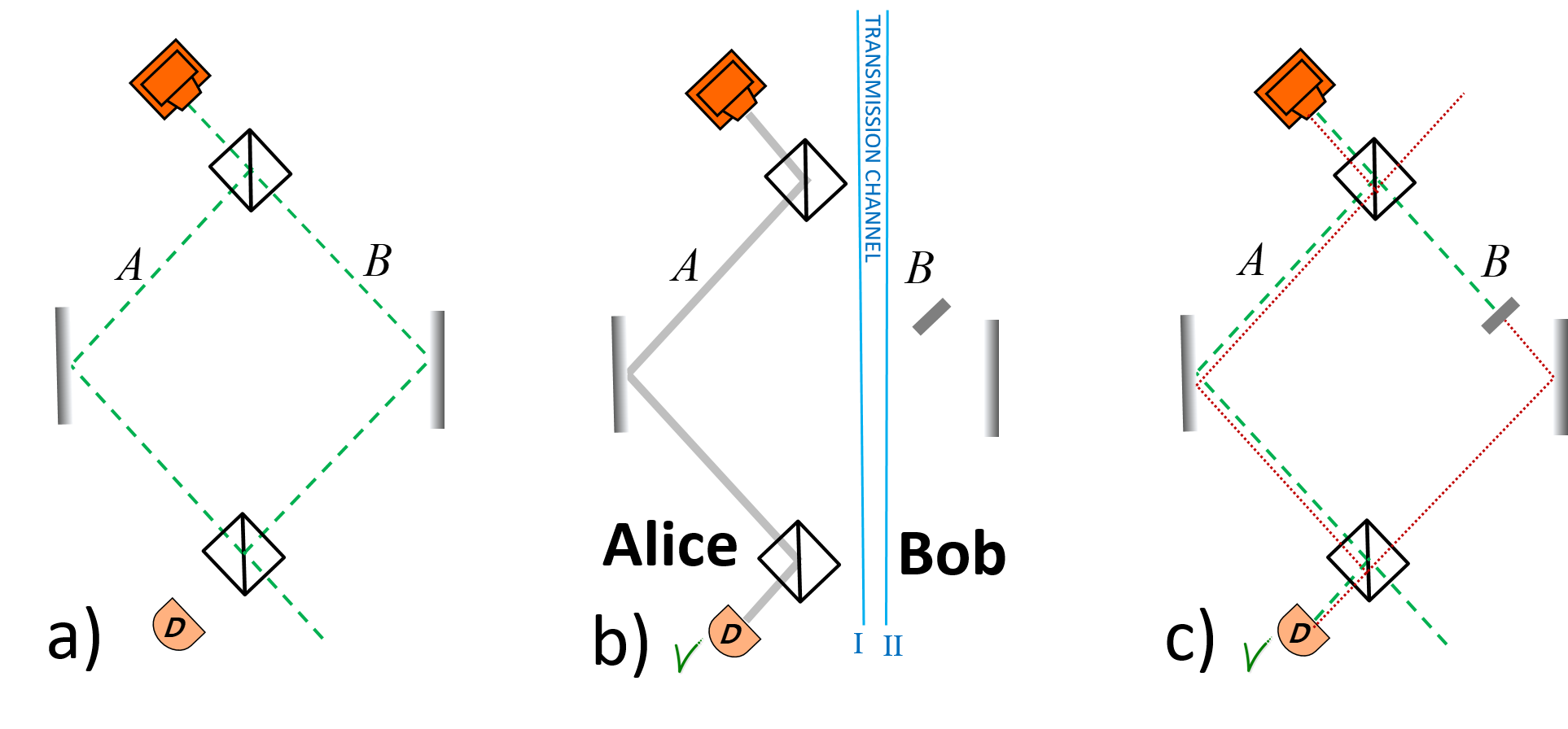}\end{center}
\caption{Counterfactual detection of the presence of the shutter. a) The interferometer is tuned in such a way that  detector $D$ never clicks if the paths are free. b) Alice  knows that Bob chose bit 1 (blocked the path) when she observes the click in $D$. Gray thick line shows the trace left by the photon. It does not present in the transmission channel. c) Forward and backward evolving wave functions of the photon. }
\label{fig:1}
\end{figure}

\section{Interaction-free measurement of the presence of a shutter}

The basic counterfactual protocol, the IFM, is shown on Fig. 1. The  photon  in tuned Mach-Zehnder interferometer cannot  reach detector $D$ when there is nothing disturbing the photons inside the interferometer. It can click  when we place an object in one arm of the interferometer. Considering everything to the left of the line I as the place of Alice, everything to the right of line II as the place of Bob, and everything between lines I and II as the transmission channel, the IFM is a counterfactual communication of a single value of  a bit. Presence of a shutter on Bob's site we define as 1 and absence as 0. For value 1, Alice sends a single photon and she has a chance to get the click in $D$. Then she knows the bit and we can also claim that no particle was in the transmission channel. One argument (which we do not accept as legitimate) is that if it would be in the channel, we would not be able to get the click in $D$. But there is also another argument which we do find decisive: after performing the protocol, no trace is left in the transmission channel, see Fig. 1b. This can be easily seen from the fact that at no point of the channel there is an overlap of the forward and backward evolving states, Fig. 1c.

Let us also demonstrate it without two-state vector formalism by analyzing a model of the trace. We consider that every arm of the interferometer is a channel with a small coupling to the photon passing through it. In our model  the state of the photon passing through a channel is not changed, but the quantum state of each channel, originally described by  $| \chi \rangle$, is modified due to the passage  of the photon:
\begin{equation}\label{eq::iaSingle}
|\chi \rangle \rightarrow|\chi^\prime \rangle \equiv \eta |\chi \rangle + \epsilon |\chi^\perp \rangle ,
\end{equation}
where $| \chi^\perp \rangle$ denotes the component of $| \chi^\prime \rangle$ which is orthogonal to $| \chi \rangle$ and its phase is chosen such that $\epsilon > 0$. The same analysis is valid if the orthogonal component appear not in the state of the physical channel, but in other degree of freedom of the photon itself \cite{Univ}.

 Since optical interferometers achieve very high fidelity, the amplitude of the orthogonal component is very small, so we can assume, $\epsilon \ll 1$. The model allows us to quantify our definition of the presence of the particle. If it leaves an orthogonal component with amplitude of  order $\epsilon$ (even if $\epsilon$ is very small), then we   say that the photon was in this channel.  If the channel remained undisturbed or it left with only high order contributions in $\epsilon$, which means, since $\epsilon \ll 1$, that the trace is arbitrarily small compared with the trace of a single particle,  we  declare that the  photon was not present in the channel.

 Let us consider the evolution of the states of the photon and  of the channels in the successful IFM protocol, Fig.~1b.
  Splitting at the first beam splitter
\begin{equation}\label{01}
|IN \rangle|\chi \rangle_A |\chi \rangle_B\rightarrow\frac{1}{\sqrt{2}}\left(|A\rangle| + |B\rangle\right)|\chi \rangle_A |\chi \rangle_B,
\end{equation}
coupling to the channels
\begin{equation}\label{02}
\rightarrow\frac{1}{\sqrt{2}}\left(|A\rangle|\chi^\prime \rangle_A |\chi \rangle_B + |B\rangle|\chi \rangle_A |\chi^\prime \rangle_B\right),
\end{equation}
collapse when the photon is not absorbed by the shutter
\begin{equation}\label{03}
{\!} \rightarrow {\!} |A\rangle |\chi^\prime \rangle{\!}_{_A}|\chi \rangle{\!}_{_B} {\!},
\end{equation}
and then collapse when detector $D$ clicks
\begin{equation}\label{03}
 {\!}\rightarrow{\!} |D\rangle |\chi^\prime \rangle{\!}_{_A}|\chi \rangle{\!}_{_B}
{\!} \simeq {\!}|D\rangle {\!}\left(|\chi \rangle{\!}_{_A} {\!} +{\!} \epsilon|\chi^\perp \rangle{\!}_{_A} {\!} \right)|\chi \rangle{\!}_{_B}.
\end{equation}
We see that the orthogonal component is present only in path $A$. There is no trace in path $B$ near the shutter, so it is a counterfactual communication of bit 1.

\section{Interaction-free measurement of the absence of a shutter}

The next ingredient of counterfactual protocols is transmitting bit 0, corresponding to the absence of the shutter. This apparently can be achieved using nested Mach-Zehnder interferometer, Fig. 2. The inner interferometer is balanced and it is tuned to destructive interference toward the final beam splitter of the external interferometer, Fig.~2a. The external interferometer has beam splitters with transmissivity 1:2 and it is  tuned  such that the photon  cannot  reach detector $D$ when arm $B$ is blocked, Fig. 2b. It can reach detector $D$  when nothing is blocked inside the interferometer. Thus, when Alice sends a single photon and it is detected in $D$, she knows that Bob did not put the shutter in arm $B$. Using classical physics approach, Alice also might claim that this was an event of counterfactual communication. The photon could not have been in arm $B$ because photons entering inner interferometer could not reach detector $D$.
\begin{figure}
\begin{center}
 \includegraphics[width=7.4cm]{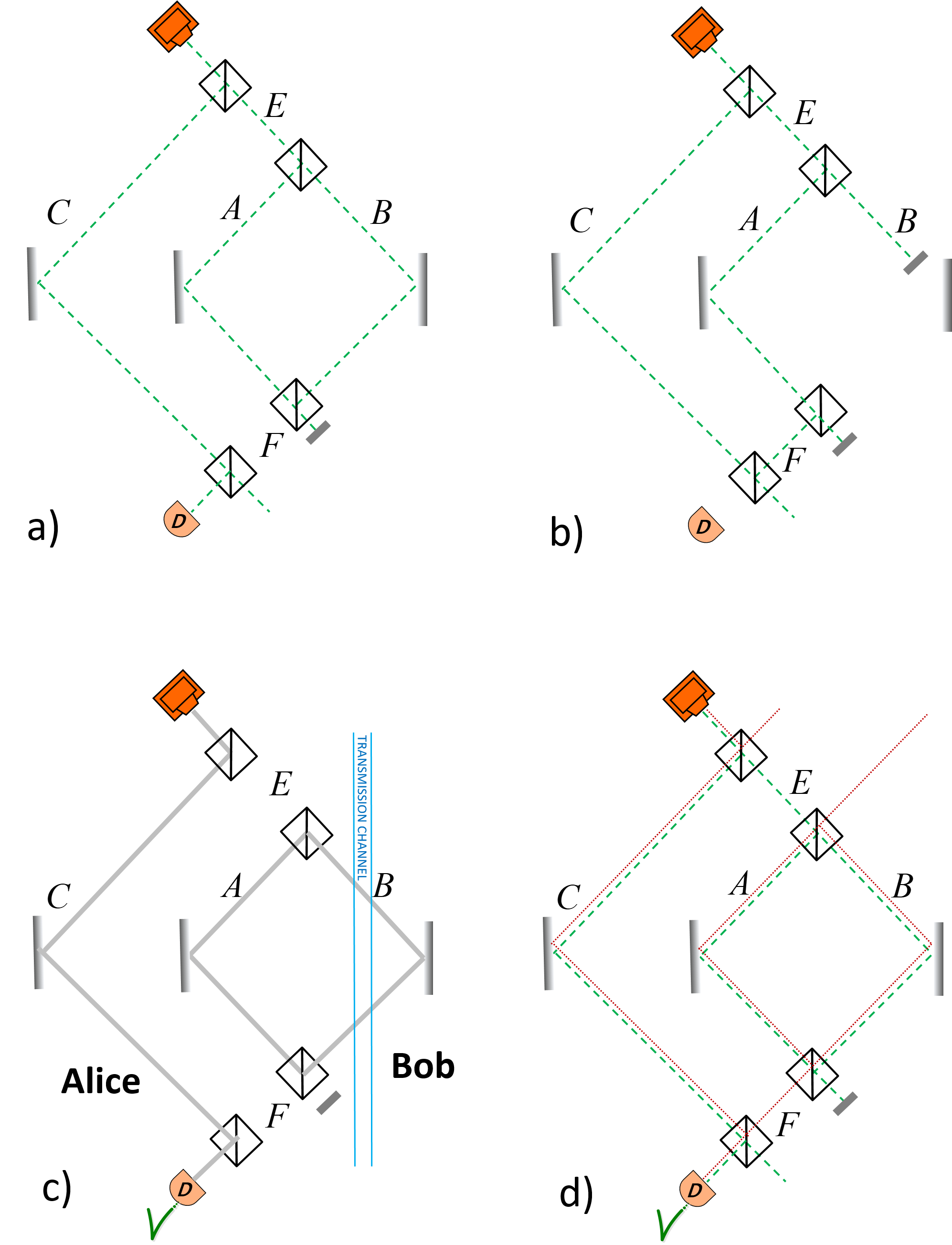}\end{center}
\caption{ Counterfactual detection of the absence of the shutter. a) The inner interferometer is tuned to destructive interference toward the continuation in the large interferometer. b) The whole interferometer is tuned such that when arm $B$ is blocked, detector $D$ cannot click. c) There is a trace in arm $C$ and inside the inner interferometer. In particular, there is a trace in the transmission channel which contradicts counterfactuality of the protocol. d) The overlap of the forward and the backward evolving waves explains the weak trace in the interferometer. }
\label{fig:2}
\end{figure}

Although the photon could not pass through $B$, it left a significant trace there, the same trace as in $C$, where everyone agrees about the presence of the photon, see Fig. 2c. Both in $C$ and in $B$ (and also in $A$) there is an overlap of the forward and backward evolving states, Fig.~2d.

Let us use also our trace model. The photon in the middle of the interferometer is in the state $\frac{1}{\sqrt{3}}\left(|A\rangle  + |B\rangle + |C\rangle  \right)$. After detection of the photon at $D$, corresponding to projection on the photon state $\frac{1}{\sqrt{3}}\left( |B\rangle -|A\rangle + |C\rangle  \right)$, the quantum state of the arms $A$, $B$ and $C$ is approximately
\footnotesize
\begin{equation}\label{33}
|\chi\rangle_A|\chi \rangle{\!}_{_B} |\chi \rangle{\!}_{_C} + \epsilon{\!}\left( |\chi\rangle{\!}_{_A}|\chi^\perp{\!} \rangle{\!}_{_B} |\chi \rangle{\!}_{_C}-|\chi^\perp{\!}\rangle{\!}_{_A}|\chi \rangle{\!}_{_B} |\chi \rangle{\!}_{_C}  +|\chi\rangle{\!}_{_A}|\chi \rangle{\!}_{_B} |\chi^\perp {\!} \rangle{\!}_{_C} {\!} \right).
\end{equation}
\normalsize
We see that  orthogonal components of order $\epsilon$ are present in arms $A$, $B$, and $C$. The protocol was supposed to find that the arm $B$ is empty without the photon being there, but the photon left a trace there.

It is of interest also to ask the question about arms $E$ and $F$: Was the photon there? Considering our trace model in all arms of the interferometer, we see that after detection of the photon at $D$, the lowest order terms with orthogonal components in arms $E$ and $F$ are
\small
\begin{equation}\label{3}
 \epsilon^2|\chi \rangle{\!}_{_C}\left(|\chi \rangle{\!}_{_F}|\chi^\perp \rangle{\!}_{_E} +|\chi^\perp \rangle{\!}_{_F}|\chi \rangle{\!}_{_E} \right)\left(|\chi\rangle{\!}_{_A}|\chi^\perp \rangle{\!}_{_B}  - |\chi\rangle^\perp{\!}_{_A}|\chi \rangle{\!}_{_B}   \right).
\end{equation}
\normalsize
 According to our definition, we need the first order in $\epsilon$ to claim the presence of the photon, so the photon was not in $E$ and $F$.

 The fact that in this protocol no first order trace is left in $E$ and $F$ allows to claim that the protocol is a counterfactual transmission of bit 0. We can modify the transmission channel of Fig. 2c. such that the boundary between Alice and Bob will pass  through $E$ and $F$ instead of $B$. Then Alice gets information that Bob's arm is empty without any particle present in the channel connecting between Alice and Bob. This, however, is not an interaction-free measurement telling us that a particular place is empty without any particle being there. The trace of the photon is left in this place.

 In counterfactual communication protocols this feature corresponds to the lack of the trace in different parts of the transmission channel for different values of the transmitted bit. Thus, by placing different boundaries of Alice's and Bob's sites  we can have counterfactual communication for one or the other value of a bit, but not for both. And, for ``counterfactual'' communication of a qubit, the trace is left in all parts of the transmission channel.

\section{Proposed modification}

To make the protocol for detecting absence of the shutter counterfactual not only according to illegitimate classical argument, but also according to the quantum ``no trace'' criterion, we propose a scheme presented in Fig.~3. It is essentially two interferometers of Fig.~2. connected by a double-sided mirror.  The inner interferometers  are balanced and tuned, as before, to destructive interference toward the path of the external interferometer, Fig.~3a.
 The external interferometer has beam splitters with transmissivity 1:4 and it is  tuned  such that when  inner interferometers have blocked arms $B$ and $B'$, we get destructive interference toward detector $D$, Fig.~3b. So again, since it is arranged  that there are only two options, either the arms $B$ and $B'$ are blocked, or the two arms are open, the click in $D$ tells us that both are open. Alice knows that shutters are absent.

 Classical argument tells us that the particle was not in arms $B$ and $B'$ since photons entering inner interferometer cannot reach Alice's detector. More importantly, the trace criteria tells us that the photon was not in arms $B$ and $B'$. We can see from  Fig.~3c that the forward and backward evolving wave functions overlap neither in $B$ nor in $B'$, and therefore, there is no trace in the transmission channel,  Fig.~3d.

 Our trace model calculations tell us the same. After detection of the photon at $D$,
 the lowest order terms with orthogonal components in arms $B$ and $B'$ are
\footnotesize
\begin{eqnarray}\label{4}\nonumber
 \epsilon^2{\!\!\!\!} \prod_{X\neq A,B,A',B'}{\!}|\chi \rangle{\!}_{_X} ~(|\chi \rangle{\!}_{_B}|\chi \rangle{\!}_{_{B'}}|\chi^\perp \rangle{\!}_{_A} |\chi^\perp\rangle{\!}_{_{A'}}   - |\chi^\perp \rangle{\!}_{_B}|\chi \rangle{\!}_{_{B'}}|\chi \rangle{\!}_{_A} |\chi^\perp\rangle{\!}_{_{A'}} \\
 - |\chi \rangle{\!}_{_B}|\chi^\perp \rangle{\!}_{_{B'}}|\chi^\perp \rangle{\!}_{_A} |\chi\rangle{\!}_{_{A'}}  +   |\chi^\perp \rangle{\!}_{_B}|\chi^\perp \rangle{\!}_{_{B'}}|\chi \rangle{\!}_{_A} |\chi\rangle{\!}_{_{A'}}  ) .~~~~
\end{eqnarray}
\normalsize
Again, since there are no orthogonal components with amplitude in the first order in $\epsilon$, the photon was not in the transmission channel. The photon passed solely through arm $C$, only there we left with the first order term, $ \epsilon |\chi^\perp \rangle{\!}_{_{C}}\prod_{X\neq C}|\chi \rangle{\!}_{_X}$.

\begin{figure}
\begin{center}
 \includegraphics[width=7.2cm]{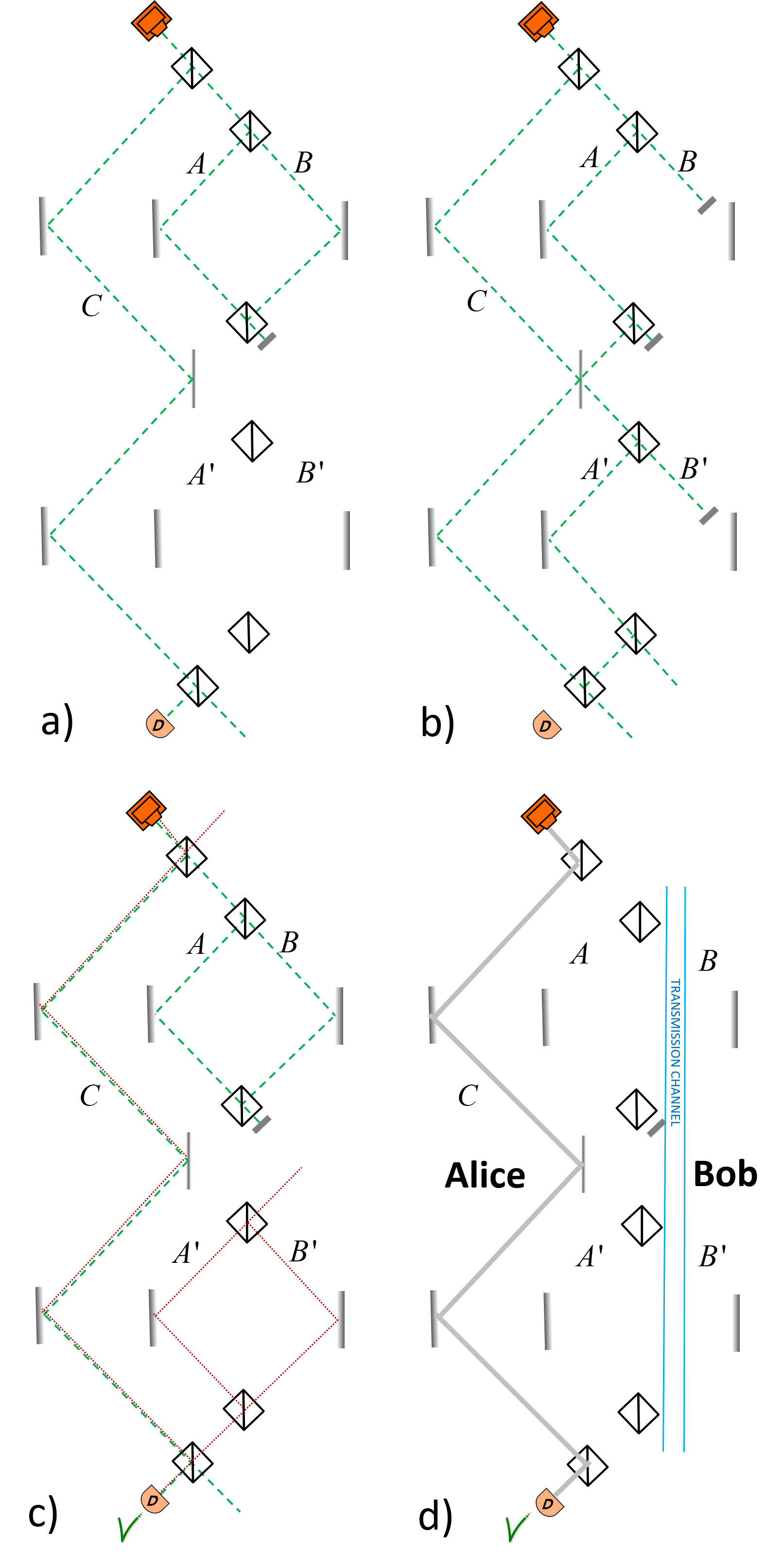}\end{center}
\caption{ Modified bit 0 counterfactual  communication. a),b) describe the tuning of the interferometer: a) shows forward evolving wave function with the shutters and b) without the shutters; the whole interferometer is tuned such that when arms $B$ and $B'$ are blocked, detector $D$ cannot  click. c) Forward and backward evolving states. d)Trace of the photon. }
\label{fig:3}
\end{figure}

Even in our improved protocol,  the trace  in the transmission channel when sending bit 0 is not exactly zero as in the case of communication of bit 1. Some  decoherence of the  photon is always present and we never get perfect destructive interference. Thus, there is a tiny leakage of the forward evolving wave toward the lower interferometer and of the backward evolving wave toward the upper interferometer, creating some overlap of the forward and the backward evolving wave functions and, therefore, some trace in the transmission channel. However, this trace, as we have shown, is of the second order in $\epsilon$. It is  much smaller than the trace of a single particle passing through the channel and thus, according to the weak trace criterion, it should be neglected.

\section{Modified counterfactual communication protocol}

The scheme for communication of bit 1 and the scheme for communication of bit 0 presented above are not the same, so we do not have yet a counterfactual communication protocols for all values of the bit. The ingenious combination of the two with help of quantum Zeno effect presented in \cite{Ho06,Salih} provides the counterfactual communication protocol. The original proposal includes the chain of external interferometers, each one with a chain of inner interferometers.  It is a very reliable communication protocol, it succeeds with probability very close to 1. The probability of the failure (loosing the photon or giving erroneous outcome) goes to zero  with  increasing the number of elements  in the chains of the interferometer.

As mentioned above, the problem is that while the case with shutters is unquestionably counterfactual, the case without shutters  is counterfactual only according to the naive classical argument: all particles passing to Bob's territory through the transmission channel could not reach Alice's detector where it was postselected. Nevertheless, during the process,  a weak trace is left in the transmission channel. It can be seen just from drawing forward and  backward evolving states, they overlap in the communication channel, see one element of the external chain in Fig.~4a. The weak trace is shown in Fig.~4b. Our trace model also shows this, the trace left in the transmission channel during the protocol is larger than the trace of a single particle passing from Alice to Bob. The calculations are shown in detail in \cite{count}.

The simple correction  method discussed above works here too. We just double each element of the chain of external interferometers  connecting them with the two-sided mirror, Fig.~4c.  When Bob places all shutters in, the protocol works as before  except for doubling the probability of losing  the photon, which is not a problem since it is very small. When Bob does not put the shutters, the communication happens  exactly as before (given ideal mirrors), but without  trace in the transmission channel, see Fig.~4d.  At least, there is no trace of the order of the trace left by a particle passing through the channel. Indeed, the weak value of local operators in the transmission channel vanish, and therefore no trace of the first order in the interaction coupling of the photon with the channel is present. Our trace model also shows this in a simple way. We get only second order contribution because photon has to leave a trace in both small chains to reach the detector.

\begin{figure}
\begin{center}
 \includegraphics[width=7.2cm]{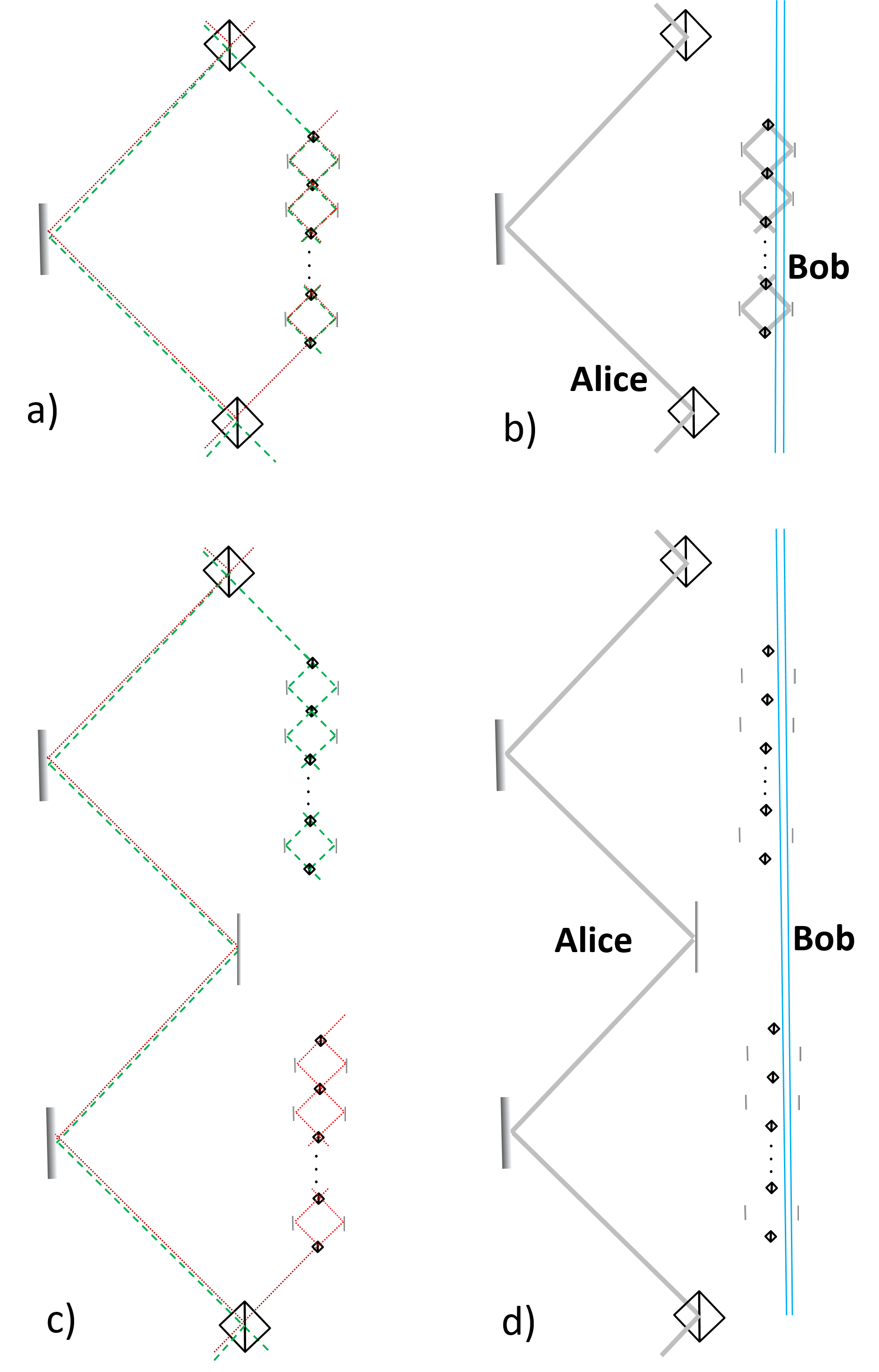}\end{center}
\caption{ a-b) One element of the chain of interferometers according to the old proposal for  counterfactual  communication. a) Forward and backward evolving states, b) the weak trace.  c-d) The same for the modified element of the chain of the counterfactual protocol. There is no trace in the communication channel. }
\label{fig:4}
\end{figure}

\section{Discussion and conclusions}

Considering the shutter as a quantum computer performing calculation of a binary function of a binary input provides a method for counterfactual computation for all possible variables. The protocol \cite{Ho06} with this simple modification achieves the task. And it is definitely a feasible task. The large interferometer  with the chain of the units of the form presented  in Fig.~4c. is not needed. Just three coupled optical cavities with two high-reflectivity beam splitter  with one of them convertible into a two-sided mirror.  Essentially the same experiment that has been performed, only the opening of the first beamsplitter happening after twice the time it was done originally. The same is true for the setup described in \cite{Salih} and other variations.

 Does it contradict the general limitation on counterfactual computation derived by Mitchison and Jozsa \cite{Mi-Jo}?  No, we do not have here a single  (counterfactual) operation of the computer. We need multiple  identical computers  or the same computer interrogated many times.

Did we found an IFM that a particular place is empty? Almost. Our method requires a promise: either two places are empty or two places blocked. It also equivalent to the promise that either a particular place is empty at two times or it is blocked at two times. It does not achieve the dual task to the original IFM \cite{IFM} of finding an object at a particular place at a particular time without particles being there.

The protocol of counterfactual communication of classical information explained above can be generalized to transmitting an arbitrary quantum state as explained in \cite{SalihQT,Li,Mycom}. It is a quantum state of multiple shutters: a superposition of the state of shutters all blocking paths $B$ of the interferometer with the state when they all are outside the interferometer.

Counterfactual transfer of a quantum state looks like  an improved version of quantum teleportation \cite{tele}: there is no need for preparation of a quantum entangled particles and nothing is transmitted between Alice and Bob, neither quantum particles,  nor classical information.  However, it does not have a practical advantage. The method requires multiple quantum channels to be build and/or multiple operations in time to be (counterfactually) performed. It takes much longer time. Also, given  ideal devices, teleportation always succeeds, while counterfactual transmission succeeds only with probability arbitrary close to 1, but not 1.

Communication without particles moving in the transmission channel, and, especially transmission of a quantum state without presence of any particle  in the transmission channel is a bizarre feature of quantum theory. It tells us that quantum theory must have  some kind of action at a distance. One of us, LV, wants to mention that there is a way to escape action at a distance  for the price of accepting existence of multiple parallel worlds \cite{myMWI}. The physical intuition that nothing can happen without causal local action can be restored by applying physical intuition to all worlds together. The tiny probability of the failure of the protocol corresponds to existence of numerous other worlds  in which the photon did pass through the channel.

Another consistent approach is not to ask where was the photon inside the interferometer. Analysis of the evolution of the forward evolving wave function (which passes through the transmission channel) explains all observable  results. Still, operational meaning of quantum particles as leaving a trace where they pass is a helpful feature describing quantum systems, especially of pre- and postselected quantum systems. It is useful and important to investigate the limits of classical description of our quantum world.

 There were several experiments  performing protocols for counterfactual communication which are not counterfactual according to the criterion of the weak trace in the transmission channel \cite{Ho06,Salih,K-Du,Chao,PNAS}. It will be of interest to repeat these experiments with the modification proposed here. Even more interesting, although  more challenging, is to experimentally compare between the weak traces left by the particle  in the  transmission channel in the original and in the modified schemes of counterfactual communication.

After submission of this manuscript, another proposal for counterfactual communication without weak traces was proposed and even demonstrated experimentally \cite{SalihPreprint}. The protocol is more complicated (it adds manipulation of polarization) and, by construction, it has a finite probability of an error.

This work has been supported in part by the Israel Science Foundation Grant No. 1311/14 and
the German-Israeli Foundation for Scientific Research and Development Grant No. I-1275-303.14.

\end{document}